\documentstyle[epsfig,wrapfig]{aipproc}

\begin{document}

\title{First Results of a Study of TeV Emission from GRBs in Milagrito}

\author{
J.~E.~McEnery$^{1}$,
R.~Atkins$^{1}$,
W.~Benbow$^{2}$,
D.~Berley$^{3,10}$,
M.L.~Chen$^{3,11}$,
D.G.~Coyne$^{2}$,
B.L.~Dingus$^{1}$,
D.E.~Dorfan$^{2}$,
R.W.~Ellsworth$^{5}$,
D.~Evans$^{3}$,
A.~Falcone$^{6}$,
L.~Fleysher$^{7}$,
R.~Fleysher$^{7}$,
G.~Gisler$^{8}$,
J.A.~Goodman$^{3}$,
T.J.~Haines$^{8}$,
C.M.~Hoffman$^{8}$,
S.~Hugenberger$^{4}$,
L.A.~Kelley$^{2}$,
I.~Leonor$^{4}$,
M.~McConnell$^{6}$,
J.F.~McCullough$^{2}$,
R.S.~Miller$^{8,6}$,
A.I.~Mincer$^{7}$,
M.F.~Morales$^{2}$,
P.~Nemethy$^{7}$,
J.M.~Ryan$^{6}$,
B.~Shen$^{9}$,
A.~Shoup$^{4}$,
C.~Sinnis$^{8}$,
A.J.~Smith$^{9}$,
G.W.~Sullivan$^{3}$,
T.~Tumer$^{9}$,
K.~Wang$^{9}$,
M.O.~Wascko$^{9}$,
S.~Westerhoff$^{2}$,
D.A.~Williams$^{2}$,
T.~Yang$^{2}$,
G.B.~Yodh$^{4}$ 
}

\address{(1) University of Utah, Salt Lake City, UT\,84112, USA \\
(2) University of California, Santa Cruz, CA\,95064, USA\\
(3) University of Maryland, College Park, MD\,20742, USA\\
(4) University of California, Irvine, CA\,92697, USA\\
(5) George Mason University, Fairfax, VA\,22030, USA\\
(6) University of New Hampshire, Durham, NH\,03824, USA\\
(7) New York University, New York, NY\,10003, USA\\
(8) Los Alamos National Laboratory, Los Alamos, NM\,87545, USA\\
(9) University of California, Riverside, CA\,92521, USA\\
(10) Permanent Address: National Science Foundation, Arlington, VA\
,22230, USA\\
(11) Now at Brookhaven National Laboratory, Upton, NY\,11973, USA}  

\maketitle

\begin{abstract}
Milagrito, a detector sensitive to $\gamma$-rays at TeV energies,
monitored the northern sky during the period February 1997 through
May 1998.  With a large field of view and high duty cycle, this
instrument was used to perform a search for TeV counterparts to
$\gamma$-ray bursts.  Within the Milagrito field of view 54
$\gamma$-ray bursts at keV energies were observed by the Burst And
Transient Satellite Experiment (BATSE) aboard the Compton Gamma-Ray
Observatory.  This paper describes the results of a preliminary
analysis to search for TeV emission correlated with BATSE detected
bursts.  Milagrito detected an excess of events coincident both
spatially and temporally with GRB 970417a, with chance probability
$2.8 \times 10^{-5}$ within the BATSE error radius. No other
significant correlations were detected. Since 54 bursts were examined
the chance probability of observing an excess with this significance
in any of these bursts is $1.5 \times 10^{-3}$.  The statistical
aspects and physical implications of this result are discussed.
\end{abstract}

\section{Observations and Analysis}

Milagro, a new type of TeV $\gamma$-ray observatory sensitive at
energies above 100 GeV, with a field of view of over one steradian and
a high duty cycle, began operation in February 1999, near Los Alamos,
NM.  A predecessor of Milagro, Milagrito~\cite{atkins99b}, operated
from February 1997 to May 1998. During this time interval, 54
$\gamma$-ray bursts (GRBs) detected by BATSE~\cite{paciesas99} were within
Milagrito's field of view (less than 45$^{\circ}$ zenith angle).


A search was conducted in the Milagrito data for an excess of events
above the cosmic-ray background coincident with each of these
$\gamma$-ray bursts. For each burst, a circular search region was
defined by the BATSE 90\% confidence interval, which incorporates both
the statistical and systematic position errors~\cite{briggs99}.  The
size of this 90\% confidence interval ranged from 4$^{\circ}$ to
26$^{\circ}$ for the 54 GRBs in the sample.  The search region was
tiled with an array of overlapping $1.6^{\circ}$ radius bins centered
on a $0.2^{\circ} \times 0.2^{\circ}$grid. This radius was derived
from the measured angular resolution of Milagrito and was selected
prior to the search.  The number of events falling within each of the
$1.6^{\circ}$ bins was tallied for the duration of the burst reported
by BATSE.  This duration is defined as the time required for BATSE to
accumulate 90\% of the $\gamma$-rays(T90). T90 ranged from 0.1 seconds
to 195 seconds for the 54 bursts examined.


The angular distribution of background events on the sky was
characterized using two hours of data surrounding each burst. This
distribution was normalized to the number of events detected by
Milagrito over the entire sky during the T90 interval ($N_{T90}$). The
resulting background data were also binned in $1.6^{\circ}$ bins
spaced $0.2^{\circ}$ apart. The Poisson probability that the excess of
events in each $1.6^{\circ}$ bin was due to a background fluctuation
was calculated and the bin with lowest probability was taken as
the candidate position of a TeV $\gamma$-ray counterpart to the BATSE
burst. The background and signal counts in this bin were used
to calculate a fluence or fluence upper limit for each burst.

\section{results}

The flux sensitivity of Milagrito to $\gamma$-ray bursts depends on
the zenith angle and duration of the burst, and on the instrument
conditions at the time.  During the lifetime of the Milagrito detector,
data were taken with three different water depths (0.9 m, 1.5 m and
2.0 m). In addition, for the period February 1997 through the end of
March 1997 a considerable amount of snow collected on the cover of the
pond. Detector simulations were used to obtain effective area as a
function of zenith angle for an assumed E$^{-2.0}$ spectrum for each
of these configurations . These were then used to calculate flux upper
limits for each burst.  Flux upper limits in the range $10^{-6} -
10^{-8}$~$\gamma$/cm$^{2}/s$ were obtained for 53 of the 54 bursts in
the sample.

\begin{wrapfigure}[17]{l}{8cm}
\vspace{-0.7cm}
\centerline{\epsfig{figure=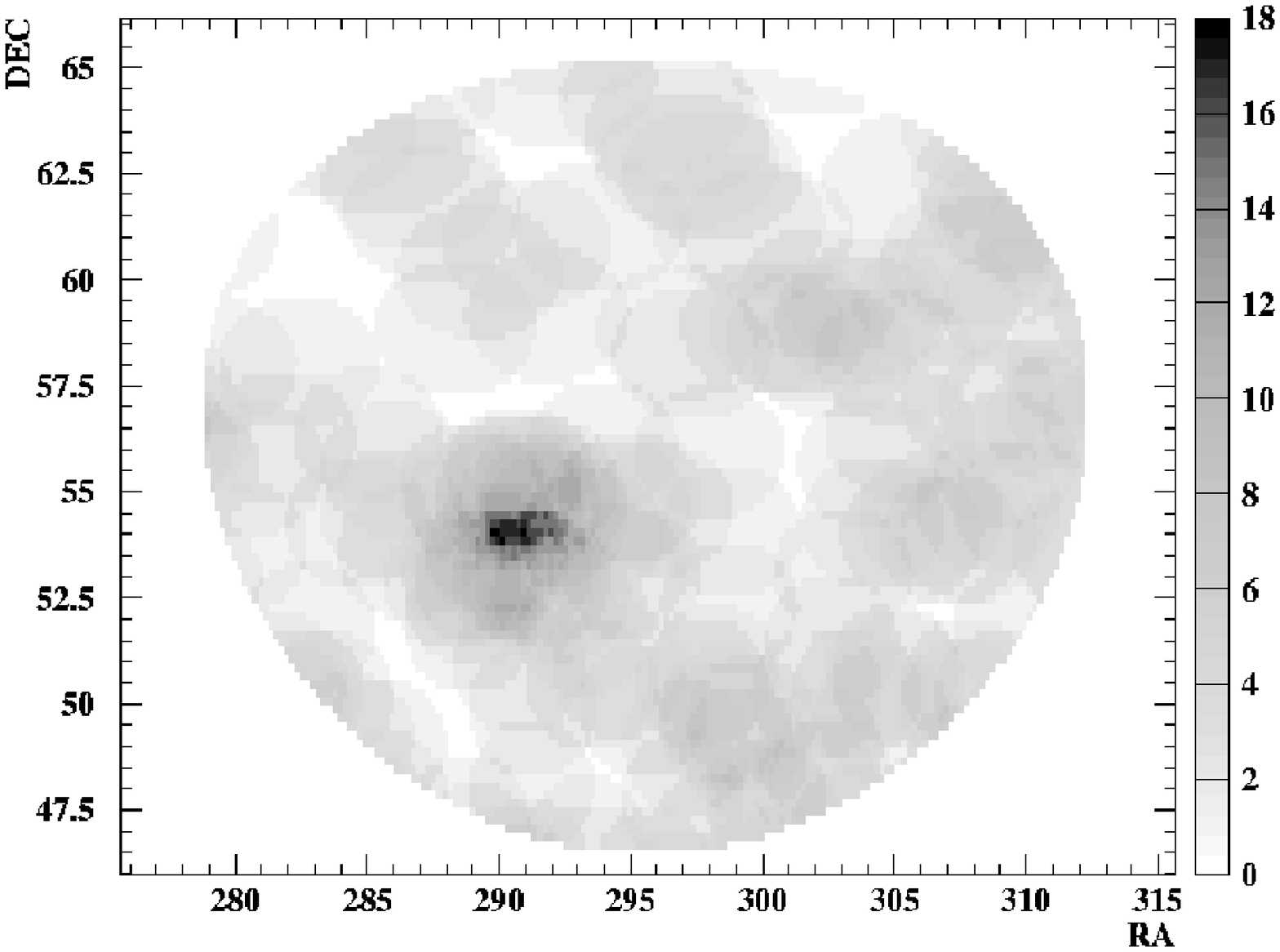,width=8cm,angle=180}}
\caption{Number of events recorded by Milagrito during T90 in the
BATSE error radius for GRB 970417a, each bin contains the number of 
events detected by Milagrito within a 1.6 degree radius.}
\label{fig:t90sky}
\end{wrapfigure}

Of the 54 bursts one, GRB 970417a, shows a substantial excess above
background in the Milagrito data.  The BATSE detection of this burst
is a weak burst with a fluence in the 50--300 keV energy range
of $1.5 \times 10^{-7}$ ergs/cm$^2$ and T90 of 7.9 seconds.  BATSE
determined the burst position to be RA~$=295.66^{\circ}$,
DEC~$=55.77^{\circ}$. The 90\% positional uncertainty was
9.4$^{\circ}$ . The $1.6^{\circ}$ radius bin with the largest excess
in the Milagrito data is centered at RA~$= 289.89^{\circ}$ and DEC~$=
54.0^{\circ}$, corresponding to a zenith angle of $21^{\circ}$.  This
position is $3.8^{\circ}$ away from the position reported by BATSE;
well within the BATSE 1-sigma position error $6.2^{\circ}$.  The
uncertainty in the position of the TeV candidate was determined by
Monte-Carlo simulations to be approximately $0.5^{\circ}$.
Figure~\ref{fig:t90sky} shows the number of counts in this search
region for the array of $1.6^{\circ}$ bins. The bin with the largest
excess has 18 events with an expected background of $3.46 \pm 0.11$.
The Poisson probability for observing a signal at least this large due
to a background fluctuation is $2.89 \times 10^{-8}$.

\begin{wrapfigure}[15]{l}{8cm}
\vspace{-0.7cm}
\centerline{\epsfig{figure=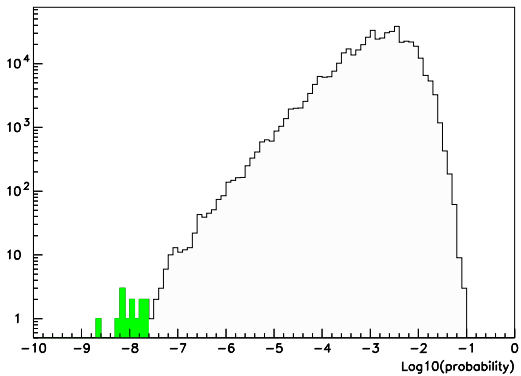,width=8cm}}
\caption{The distribution of minimum probabilities for the ensemble of
simulated data-sets for GRB 970417a.}
\label{fig:simprob}
\end{wrapfigure}

\begin{figure}[tbh!]
\vspace{-0.4in}
\centerline{\epsfig{figure=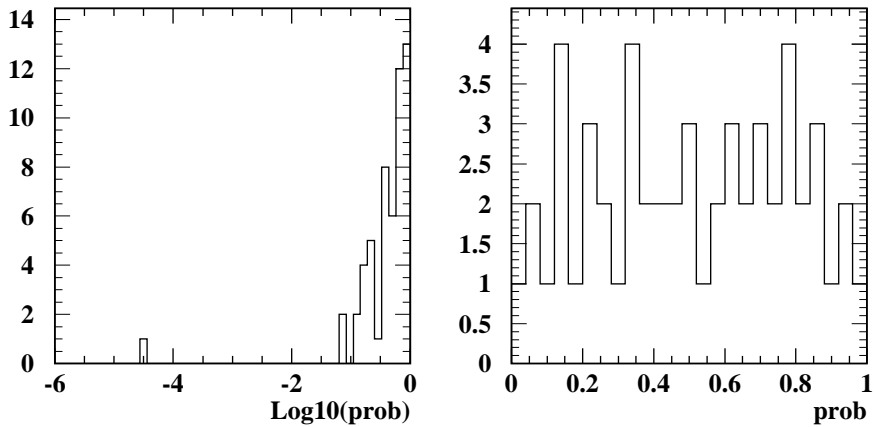,width=13.5cm}}
\caption{The distribution of probabilities, corrected for the size of the 
search region for the 54 GRBs in the sample, both plots show the same
data with a linear and logarithmic scale for the x-axis}
\label{fig:finalprob}
\end{figure}

To obtain the significance of this result one must account for the
size of the search region. The probability of obtaining the observed
significance anywhere within the entire search region was determined
by Monte Carlo simulations.  A set of simulated signal maps was made
by randomly drawing $N_{T90}$ events from the background distribution.
Each map was searched, as before, for a significant excess within the
search region defined by BATSE. The probability of the observation in
the actual data being due to a fluctuation in the background, after
accounting for the size of the search region, is given by the ratio of
the number of simulated data sets with probability less than that
observed for the actual data to the total number of simulated data
sets. The distribution of the probabilities for $4.65 \times 10^{6}$
simulated data sets is shown in figure~\ref{fig:simprob}; thirteen of
which had Poisson probability less than $2.89 \times 10^{-8}$. We
therefore find that the chance probability of such a detection within
the entire $9.4^{\circ}$ search region for GRB 970417a to be $2.8
\times 10^{-5}$.  The probabilities for each of the other 53 bursts in
the sample were obtained using the same method, the distribution of
these probabilities, after correcting for the size of the search
region, is shown in figure~\ref{fig:finalprob}. The histogram on the
left, plotted on a log-linear scale, illustrates the significance of
the excess for GRB 970417a relative to the rest of the sample. The
histogram on the right of this figure, plotted on a linear scale is
flat, as expected. 54 bursts were examined. Therefore the chance
probability of observing such a significant excess due to fluctuations
in the background for any of these bursts is $1.5\times 10^{-3}$.

\begin{figure}[tbh!]
\centerline{\epsfig{figure=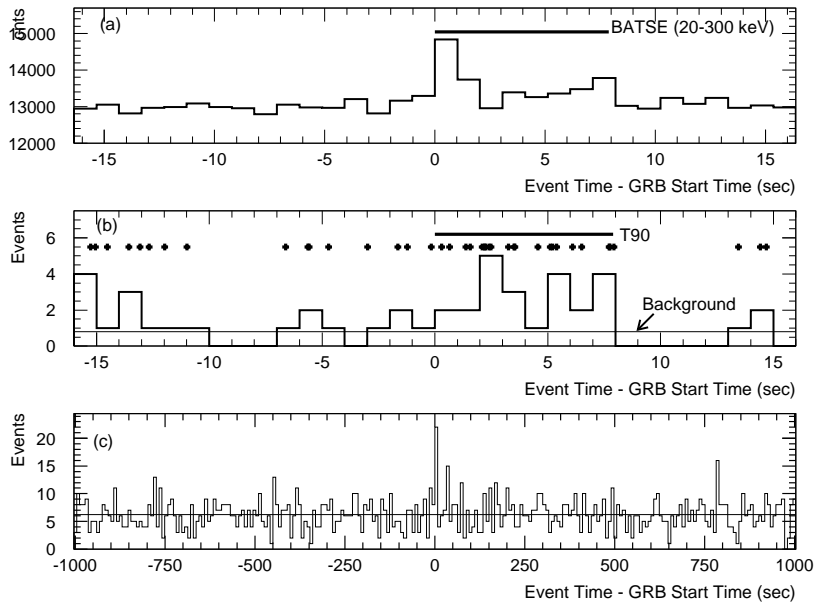,width=12cm}}
\caption{GRB 970417a: (a) The BATSE lightcurve, (b) Milagrito data
within a 2.2$^{\circ}$ radius of GRB 970417a integrated in 1 second
bins, the crosses indicate the arrival times of the events and (c)
integrated in bins of 7.9 seconds (T90) for 2000 seconds}
\label{fig:lc}
\end{figure}

Although the initial search was limited to T90, for GRB 970417a longer
time intervals were also examined. To allow for the positional
uncertainty of the excess observed by Milagrito, the radius of the
search bin was increased to $2.2^{\circ}$ for this search. A search
for TeV $\gamma$--rays integrated over long time intervals of one hour,
two hours and a day after the GRB start time did not show any
significant excess.  Lightcurves where the data are binned in
intervals of one second and of T90 (7.9 s) are shown in
figure~\ref{fig:lc}. A preliminary analysis reveals no statistically 
compelling evidence for TeV afterflares.


\vspace{-0.2in}

\section{Discussion}





\begin{figure}[tbh!]
\centerline{\epsfig{figure=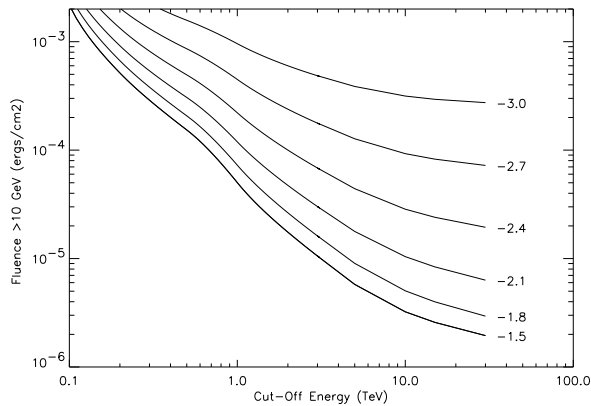,width=8cm}}
\caption{Implied fluences of this candidate for a range of assumed power-law  spectra and high energy cutoffs (preliminary)}
\label{fig:ergs}
\end{figure}

If the observed excess of events in Milagrito is indeed associated
with GRB 970417a then it represents the highest energy photons yet
detected from a GRB in coincidence with the sub-MeV emission.  The
following discussion assumes that the excess observed by Milagrito was
due to TeV $\gamma$-rays from GRB 970417a. The TeV spectrum and
maximum energy of emission is difficult to determine from Milagrito
data~\cite{atkins99b}.  Monte Carlo simulations of $\gamma$-ray
initiated air showers show that the effective area increases smoothly
with energy, making the definition of an energy threshold
ambiguous. Figure~\ref{fig:ergs} shows the implied fluence of this
observation as a function of upper cutoff energy for a range of
power-law input spectra.

Some information about the energies of the events observed for GRB
970417a can be obtained by considering the response of the summed
untriggered counting rate of the individual PMTs in Milagrito.
Detector simulations of the effect on PMT counting rates of
$\gamma$-ray induced air-showers indicate that these rates are more
sensitive than the standard shower data at energies below a few
hundred GeV, but are only sensitive to very large
fluxes~\cite{atkins99b}. No excess was observed in these rates, which
implies that the air-showers detected by Milagrito were probably due
to $\gamma$-rays at energies above several hundred GeV.


High energy $\gamma$-rays from sources at cosmological distances will
be absorbed via electron-positron pair production with infrared
photons in the intergalactic medium. Several studies find that the
opacity due to pair production for above 200 GeV $\gamma$-rays exceeds
one for redshifts larger than 0.3~\cite{salamon98,primack99}.  
Thus, if Milagrito has indeed detected high energy photons from 
GRB 990417a, it must be from a relatively nearby object.

\section{Conclusion}

An excess of events with chance probability $2.8 \times 10^{-5}$
coincident both spatially and temporally with the BATSE emission for
GRB 970417a was observed by Milagrito.  The chance probability that an
excess of this significance would be observed from the entire sample
of 54 bursts is $1.5 \times 10^{-3}$. The spectrum must extend with no
cutoff to at least a few hundred GeV.  The inferred TeV fluence from
this result at least an order of magnitude greater than the sub-MeV
fluence and the emission extends to at least several hundred GeV.

If the observed excess from GRB 970417a is not a fluctuation of the
background, then a new class of $\gamma$-ray bursts bright at
TeV energies may have been observed.  A search for other coincidences with
BATSE, to verify this result, will be continued with the current
instrument, Milagro, which has increased sensitivity to TeV
$\gamma$-ray bursts.



\acknowledgments

This research was supported in part by the National Science Foundation,
the U.S. Department of Energy Office of High Energy Physics,
the U.S. Department of Energy Office of Nuclear Physics,
Los Alamos National Laboratory, the University
of California, the Institute of Geophysics and Planetary Physics,
The Research Corporation, and the California Space Institute.


\begin{thebibliography}{99}
\bibitem{paciesas99} W. S. Paciesas et al., (Astro-Ph-9903205) (1999)
\bibitem{briggs99} M. S. Briggs et al., {\it Astrophys. J. Supp.} {\bf 122(2)}, 503 (1999)
\bibitem{salamon98} M.H. Salamon and F. W. Stecker, {\it Astrophys. J.} {\bf 493}, 547 (1998).
\bibitem{primack99} J. R. Primack et al, {\it Astroparticle Physics} {\bf 11}, 93 (1999).
\bibitem{atkins99b} R. Atkins et al., {\it Nucl. Inst. and Methods} (1999) (submitted).

\end{thebibliography}
\end{document}